\documentclass{article} % For LaTeX2e
\usepackage{iclr2027_conference,times}

\usepackage{amsmath,amsfonts,bm}

\def\eqref#1{equation~\ref{#1}}
\def\1{\bm{1}}

\DeclareMathAlphabet{\mathsfit}{\encodingdefault}{\sfdefault}{m}{sl}
\SetMathAlphabet{\mathsfit}{bold}{\encodingdefault}{\sfdefault}{bx}{n}

\usepackage{booktabs}
\usepackage{enumitem}
\usepackage{graphicx}
\usepackage{float}
\usepackage{wrapfig}
\usepackage{needspace}
\usepackage{placeins}
\usepackage{hyperref}
\usepackage{url}
\usepackage{fontawesome5}
\definecolor{cornflowerblue}{RGB}{100,149,237}

\title{Long-Term Memory-Guided Enhancement for Target Perception in Audio-Language Models}

\author{%
Zhenhong Zhou\textsuperscript{1}\thanks{Equal contribution.},\enspace
Xuanyue Zhao\textsuperscript{1$\ast$},\enspace
Youji Liu\textsuperscript{2},\enspace
Yuanhe Zhang\textsuperscript{2} \\
\textbf{Xiaoyu Ma\textsuperscript{1},\enspace
Lianyu Hu\textsuperscript{1}\thanks{Corresponding authors.},\enspace
Yang Liu\textsuperscript{1$\dagger$}} \\
\textnormal{\textsuperscript{1}Nanyang Technological University\quad
\textsuperscript{2}Beijing University of Posts and Telecommunications} \\
\multicolumn{1}{c}{\textnormal{\faEnvelope[regular]\enspace\texttt{zhenhong001@e.ntu.edu.sg}}}
}

\newif\ifshowdrafttodos
\showdrafttodosfalse

\iclrfinalcopy % Show authors and omit review line numbers in the preprint.
\begin{document}

\maketitle
\lhead{Preprint}

\begin{abstract}
Audio large language models (ALLMs) can reason about the content of audio recordings to perform complex tasks.
However, these capabilities usually collapse in real-world environments when background noise and competing sources mix the target sound.
Inspired by long-term memory in human listening~\citep{weinberger2004specific}, we propose Long-Term Memory-Guided Audio Enhancement (LTM-AE) to improve selective target perception by refining the audio representations of ALLMs without training.
LTM-AE extracts representations in hidden states from separate clean reference recordings as long-term memory for each category, guiding enhancement toward a user-specified listening target.
We reconstruct incoming audio tokens in the selected category long-term memory and interpolate the reconstructions with the original tokens before language backbone decoding.
This interpolation controls the influence of stored auditory experience while keeping all ALLM parameters fixed.
Diagnostic readouts across twenty sound categories and three ALLMs show that LTM-AE strengthens responses to a specified target amid three interfering sources.
Averaged over constrained and free-form classification, accuracy gains over raw mixtures range from 29.53 to 46.15 percentage points across multiple open source models.
For speech content recovery, LTM-AE with an additional learned token-level gate reduces \texttt{Qwen2-Audio}'s word error rate from 23.07\% to 14.77\%.
This work takes an initial step toward using principles of human long-term memory to enhance ALLMs for real-world listening.
Our code is available at~\url{https://github.com/aynlp/ltm-audio-code}
\end{abstract}

\section{Introduction}
\label{sec:introduction}

Audio large language models (ALLMs) bridge text and audio inputs, enabling language models to handle complex involving tasks the audio modality~\citep{chu2024qwen2, ding2025kimi, wu2025step}.
Real-world use brings ALLMs into diverse and changing acoustic environments that may differ quite from their training stage~\citep{li2026audiotrust,iyer2026scenebench}.
In these environments, background noise and competing sources mix even overwhelm the target audio channel~\citep{wang2018supervised,yin2026can}.
A model that can reason about clean audio may then fail to perceive the same target within a mixture~\citep{hou2025evaluating,lin2026echodistill}.
Human listening suggests that prior auditory experience can help address this gap between capability and perception under interference~\citep{agus2010rapid}.
Learning recurring sound structure also helps listeners distinguish individual sources within mixtures~\citep{munte2001superior, woods2018schema}.
These findings motivate using prior auditory experience to guide ALLMs toward a predefined listening target.

Audio enhancement has long addressed interference through statistical signal estimation and learned signal reconstruction~\citep{ephraim1984speech, luo2019conv, hu2020dccrn}.
Text descriptions are proven to guide waveform separation by specifying the sound to recover from a mixture~\citep{kilgour2022text, liu2022separate, liu2024separate, ma2024clapsep}.
For ALLMs, \citet{zhang2026rsa} have highlighted environmental interference as a major challenge, showing that competing sources can substantially impair both perception and reasoning.
In response, recent methods enhance ALLM perception by recovering a target speaker's speech~\citep{chen2026av, godiva2026noise} or selecting between speech and non-speech components~\citep{yin2026can}.
Denoising improves ALLM robustness to added noise, with the clean recording defining the content to preserve~\citep{xiong2025thinking, zhang2026see}.
However, ALLMs still struggle to perceive a specified target when multiple stronger sources compete in the same audio, especially when the target is a non-speech sound.
ALLMs have to overcome these perceptual limitations to enable broader applications in real-world scenarios such as embodied intelligence and robotics.

\begin{figure*}[t]
    \centering
    \vspace{-18pt}
    \includegraphics[width=\textwidth]{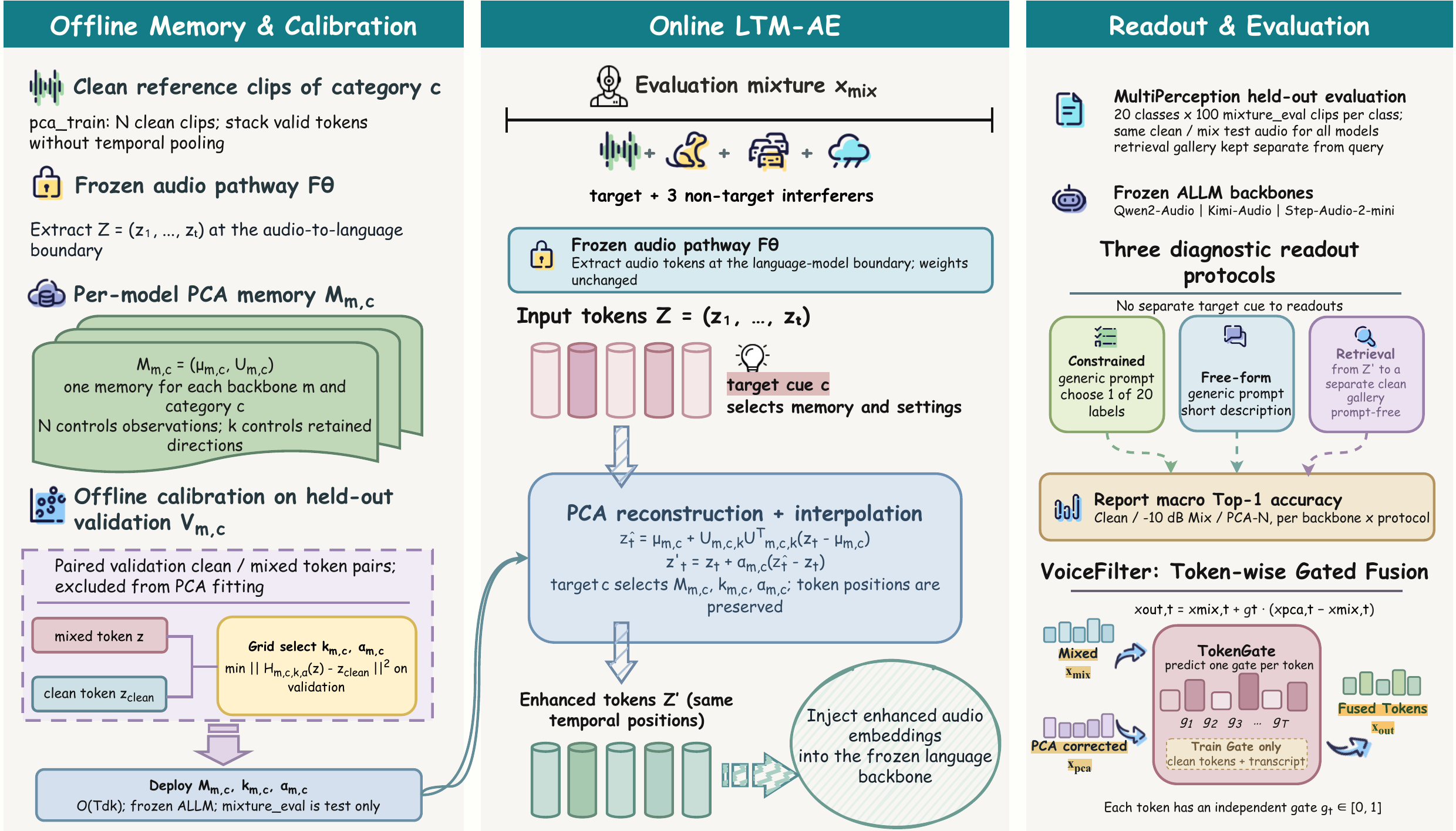}
    \caption{
Overview of LTM-AE.
Clean audio hidden states form category long-term memory, while paired validation recordings calibrate the retained rank and interpolation weight.
For an incoming mixture, the specified target category selects a memory to reconstruct audio tokens and interpolate them with the original tokens before language backbone decoding, without training the ALLM.
The speech transcription branch adds a token-level gate to adjust this guidance at each position.
}
    \label{fig:overview}
    \vspace{-13.5pt}
\end{figure*}

We propose Long-Term Memory-Guided Audio Enhancement (LTM-AE) for selective target perception through category long-term memory without training the ALLM.
The memory retains category structure from separate clean reference recordings, while the current recording determines how that structure enters the reconstruction of each audio token.
To build this memory, we extract hidden states at the interface between the audio pathway and the language backbone.
We store their mean and dominant directions of variation for reuse across later inputs.
At inference, the user-specified target category selects the memory, and the incoming tokens determine their reconstruction coefficients within it.
Interpolation with the original tokens then controls the strength of the memory's influence on the current audio representation.
Because reconstruction may discard useful target detail, we calibrate the interpolation weight together with the number of retained directions.
The calibration minimizes token reconstruction error on paired clean and mixed validation recordings, tying the enhancement to its ability to approximate clean target representations.
The language backbone decodes the resulting tokens at their original temporal positions without a separate target cue, and all ALLM parameters remain fixed.

To approximate the interference encountered in real-world acoustic environments, we synthesize thousands of audio mixtures in which a target competes with three other sound sources.
The resulting MultiPerception benchmark spans twenty target categories, including diverse non-speech sounds.
To assess the applicability of LTM-AE across models, we evaluate three open-source ALLMs: \texttt{Qwen2-Audio}, \texttt{Kimi-Audio}, and \texttt{Step-Audio-2-mini}~\citep{chu2024qwen2,ding2025kimi,wu2025step}.
With the target category specified, LTM-AE improves both constrained and free-form classification readouts across all three models without training them.
For example, \texttt{Qwen2-Audio}'s constrained classification accuracy rises from 9.60\% to 47.20\%, while its prompt-free retrieval accuracy increases from 8.25\% to 86.75\%.
Cross-model analysis shows mean gains of 29.53--46.15 percentage points across the two classification protocols.
Category-level analysis further shows retrieval gains for every category in all three ALLMs, extending the benefit to diverse non-speech targets.
We further evaluate speech transcription to test recovery of words not specified by the target category.
LTM-AE with an additional learned token-level gate reduces \texttt{Qwen2-Audio}'s word error rate from 23.07\% to 14.77\%.

Together, the method and results demonstrate how long-term auditory experience can guide selective target perception in ALLMs within complex mixtures.
We summarize our contributions as follows.(1) We introduce an approach inspired by human long-term memory that retains auditory experience for reuse in ALLMs.
We extract category long-term memory from clean audio hidden states by retaining their mean and dominant directions of variation.
(2) We propose LTM-AE to use this long-term memory for selective target perception in complex audio mixtures.
Given a specified target category, LTM-AE refines incoming audio embeddings through calibrated reconstruction and interpolation without training the ALLM.
(3) We conduct extensive experiments across three ALLMs and twenty sound categories, showing gains in both language outputs and audio representations under interference from multiple sources.
With a learned gate, LTM-AE further improves transcription of the selected target's content.

\vspace{-9pt}
\section{Related Works}
\vspace{-6pt}

\subsection{Memory in Humans and Language Models}
\vspace{-3pt}
Early studies of amnesia showed that skill learning can remain intact despite impaired recollection.
Work on priming distinguished perceptual memory from conscious remembering~\citep{tulving1990priming}.
Studies of the medial temporal lobe identified a system supporting memory for facts and events~\citep{squire1991medial}, while research on the neostriatum established a distinct system for gradual habit learning~\citep{knowlton1996neostriatal}.
Research on long-term memory examined how consolidation stabilizes newly acquired information over time~\citep{mcgaugh2000memory}.
Studies of structural plasticity connected lasting memory to changes in synaptic structure~\citep{lamprecht2004structural}.
Auditory research identified specific long-term memory traces in primary auditory cortex~\citep{weinberger2004specific}, while work on recent and remote memories examined how their organization changes with time~\citep{frankland2005organization}.
Later auditory studies investigated memory for unfamiliar sounds~\citep{agus2010rapid}, learned schemas for source segregation~\citep{woods2018schema}, and memory of recurring background noise~\citep{hicks2024noise}.
In text-based large language models (LLMs), early memory mechanisms supported generation through knowledge retrieval~\citep{lewis2020retrieval}, reuse of past representations~\citep{wu2022memorizing}, and retrieval from large text corpora~\citep{borgeaud2022improving}.
Later work retained experiences to guide agent behavior~\citep{park2023generative}, reasoning across trials~\citep{shinn2023reflexion}, and long-term conversational interaction~\citep{zhong2024memorybank}.
\vspace{-3pt}
\subsection{Denoising and Robustness in Audio Language Models}
\vspace{-3pt}
Early denoising approaches used spectral subtraction~\citep{boll1979suppression}, statistical spectral estimation~\citep{ephraim1984speech}, and wavelet shrinkage~\citep{donoho1995noising}.
Neural approaches subsequently introduced regression for speech enhancement~\citep{xu2014regression}, adversarial waveform generation~\citep{pascual2017segan}, and diffusion-based restoration~\citep{richter2023speech}.
For audio large language models, studies examined acoustic variation in spoken question answering~\citep{cui2025voxeval}, adaptation to noisy speech domains~\citep{wang2025self}, and reasoning with audio processing tools~\citep{xiong2025thinking}.
Evaluations covered task-relevant events in complex scenes~\citep{iyer2026scenebench}, interference between speech and environmental sounds~\citep{yin2026can}, and robustness in realistic acoustic scenarios~\citep{hu2026vcb}.
Further evaluation examined how environmental soundscapes affect perception and reasoning, including the effects of applying conventional denoisers before ALLM processing~\citep{zhang2026rsa}.
Representation denoising subsequently addressed interference within internal audio embeddings~\citep{zhang2026see}, while task-aware waveform enhancement selected audio components relevant to the requested task~\citep{yin2026focus}.
Other approaches aligned noisy-input responses with clean-audio supervision through self-distillation~\citep{lin2026echodistill}, connected perception with reasoning during model training~\citep{wang2026listen}, and enhanced speech tokens using audio-visual information~\citep{godiva2026noise}.

\vspace{-6pt}
\section{Long-Term Memory-Guided Audio Enhancement}
\label{sec:method}

To enhance target perception in complex audio mixtures, we introduce LTM-AE, which integrates long-term auditory experience into the representation of the current input.
We encode this experience in category long-term memory extracted from clean audio hidden states, guiding the refinement of incoming audio tokens.
We then interpolate the reconstructed tokens with the original tokens before language backbone decoding, so LTM-AE refines the audio representations of ALLMs without training.
Memory construction and calibration are performed offline, while each incoming recording is processed using the saved long-term memory and settings.

\subsection{Problem Setup}
\label{sec:memory_setting}

Given a user-specified category $c$, we consider a mixture $x$ with multiple stronger sources obscuring the target sound.
Our goal is to help the ALLM perceive this target by refining the audio representation that enters its language backbone.
We use $c$ as a listening cue to select the corresponding memory $\mathcal{M}_c$.
Let $F_{\theta}$ denote the mapping from input audio to continuous audio embeddings in the language model's input space.
We refer to these continuous embeddings as audio tokens.
For input $x$, this mapping produces a sequence $Z=F_{\theta}(x)=(z_1,\ldots,z_T)$ with $z_t\in\mathbb{R}^{d}$.
We reconstruct these embeddings with $\mathcal{M}_c$ and interpolate the reconstructions with $Z$ to obtain $Z'=(z'_1,\ldots,z'_T)$.
The refined embeddings replace $Z$ at the corresponding positions in the language model's input sequence.
The language backbone receives no separate target cue.
The number and order of audio embeddings remain unchanged, as do the text token embeddings.
We construct memory separately in each ALLM's input embedding space, with the representations used by each model detailed in Appendix~\ref{app:audio_interfaces}.
This places the retained auditory experience and the incoming mixture tokens in the same coordinates used by that model for decoding.

\subsection{Retaining Long-Term Auditory Experience}
\label{sec:memory_construction}

For each category $c$, we compute audio embeddings for $N$ clean reference clips disjoint from the evaluation recordings using the same mapping $F_{\theta}$.
We stack all valid tokens into $X_c\in\mathbb{R}^{M_c\times d}$, where $M_c$ is the total number of tokens across the reference clips.
Each token contributes a separate observation, retaining variation across temporal positions as well as across reference recordings.
We summarize these observations with their mean and dominant directions of variation~\citep{jolliffe2016principal}.
For row $X_{c,i}$ of the reference matrix, we define
\begin{equation}
\begin{aligned}
\mu_c&=\frac{1}{M_c}\sum_{i=1}^{M_c}X_{c,i}^{\top},\qquad
\widetilde X_c&=X_c-\mathbf{1}\mu_c^{\top},\qquad
\mathcal{M}_c=(\mu_c,U_c),
\end{aligned}
\label{eq:memory}
\end{equation}
where $\mathbf{1}$ is an $M_c$-dimensional vector of ones and the columns of $U_c$ are orthonormal directions that capture the dominant variation in $\widetilde X_c$.
The mean describes the category's average representation, while the orthonormal directions capture variation around that mean.
Together, they encode both a common reference point and the ways clean examples vary within the category.
For a retained rank $k$, these directions minimize the reconstruction error of the centered clean tokens,
\begin{equation}
U_{c,k}\in\underset{U^{\top}U=I_k}{\operatorname{argmin}}
\left\|\widetilde X_c-\widetilde X_cUU^{\top}\right\|_F^2,
\label{eq:memory_basis}
\end{equation}
where $U\in\mathbb{R}^{d\times k}$ and $I_k$ is the $k$-dimensional identity matrix.
We obtain these directions using matrix decomposition methods such as randomized principal component analysis (PCA)~\citep{halko2011finding} and full singular value decomposition (SVD).
Appendix~\ref{app:audio_interfaces} describes the details.
The reference count $N$ sets the number of clean clips used to construct the memory, while $k$ sets the number of retained directions.
We save $\mu_c$ and $U_c$ for reuse across input recordings.
The directions are ordered by the variation they capture, so different retained ranks use prefixes of the same matrix.

\subsection{Refining Audio Tokens with Category Long-Term Memory}
\label{sec:memory_enhancement}

For each incoming token $z_t$, we subtract the stored mean representation $\mu_c$ and compute its coefficients along the retained directions $U_{c,k}$.
We then map these coefficients back to the audio embedding space and interpolate the resulting representation with the original token,
\begin{equation}
\begin{gathered}
a_t=U_{c,k}^{\top}(z_t-\mu_c),\qquad
\widehat z_t=\mu_c+U_{c,k}a_t,\\
z'_t=z_t+\alpha_c(\widehat z_t-z_t),
\end{gathered}
\label{eq:memory_enhancement}
\end{equation}
where $U_{c,k}$ contains the first $k$ columns of $U_c$ and $\alpha_c\in[0,1]$ controls the strength of long-term memory guidance.
All tokens share the selected category long-term memory, while each incoming token determines its own reconstruction coefficients.
The coefficients $a_t$ carry information from the current recording into this shared representation, allowing the same long-term memory to produce a different correction at each position.

At $\alpha_c=0$, the ALLM receives the original tokens, while $\alpha_c=1$ uses $\widehat z_t$ directly.
Let $P_c=U_{c,k}U_{c,k}^{\top}$ and let $I$ denote the identity matrix.
We can rewrite the enhanced token as
\begin{equation}
z'_t=\mu_c+P_c(z_t-\mu_c)
 +(1-\alpha_c)(I-P_c)(z_t-\mu_c).
\label{eq:residual_shrinkage}
\end{equation}
The component represented by the retained directions is preserved, while the orthogonal residual is scaled by $1-\alpha_c$.
The retained rank therefore determines which variation is preserved, and the interpolation weight controls how strongly the remaining component is attenuated.
This attenuation may remove useful detail that the memory does not represent.
Appendix~\ref{app:memory_effect} discusses this relationship and its effect on the error relative to a clean target token.

\subsection{Calibrating Long-Term Memory Guidance}
\label{sec:memory_calibration}

To preserve target information during enhancement, we further calibrate the retained rank $k$ and interpolation weight $\alpha_c$ by comparing enhanced mixture tokens with audio embeddings from their clean targets.
For calibration, we use a separate validation set of paired mixtures and clean targets to reduce overfitting to the recordings used to construct memory.
For each category, let $\mathcal{V}_c$ contain the aligned pairs $(z,z^{\mathrm{clean}})$ of mixture and clean target tokens.
Writing $H_{c,k,\alpha}$ for the enhancement in Eq.~(\ref{eq:memory_enhancement}), we select
\begin{equation}
(k_c,\alpha_c)=\underset{(k,\alpha)\in\mathcal{G}}{\operatorname{argmin}}\;
\frac{1}{|\mathcal{V}_c|d}
\sum_{(z,z^{\mathrm{clean}})\in\mathcal{V}_c}
\left\|H_{c,k,\alpha}(z)-z^{\mathrm{clean}}\right\|_2^2,
\label{eq:memory_selection}
\end{equation}
where $\mathcal{G}$ is a finite set of candidate ranks and interpolation weights.
Each candidate is evaluated on the same aligned token pairs, and the objective averages the squared discrepancy over tokens and embedding coordinates.
We repeat this selection for each reference count $N$.
Calibration selects the reconstruction settings while keeping the category mean and directions fixed, and LTM-AE does not train any ALLM parameters.
We save the selected settings with the corresponding long-term memory. At inference, enhancement requires these saved quantities and the incoming audio tokens.

Overall, LTM-AE first builds category long-term memory from clean audio embeddings and calibrates the retained rank and interpolation weight on a separate validation set.
For the mixture, we extract its audio embeddings and select the memory corresponding to the specified target category.
Using the calibrated settings, we interpolate the representations guided by long-term memory with the original embeddings and pass the refined sequence to the language backbone for decoding.
This process uses long-term auditory experience to guide target perception in complex mixtures without training the ALLM.

\section{Experiments}
\label{sec:experiments}

In this section, we evaluate the effectiveness of LTM-AE in helping ALLMs perceive a specified target under acoustic interference.
Section~\ref{sec:final20_results} presents diagnostic readouts of selective target perception through language outputs and audio representations.
We then analyze the gains across ALLMs and sound categories in Section~\ref{sec:enhancement_gain_analysis}.
Finally, Section~\ref{sec:speech_transcription} explores an extension to speech transcription with a learned token-level gate.

\paragraph{Experimental Setup}
\label{sec:experimental_setup}

We use MultiPerception to evaluate target perception under acoustic interference.
The benchmark comprises twenty target sound categories spanning musical instruments, human speech, animal vocalizations, and environmental events.
Each mixture combines a target with three interferers.
Appendix~\ref{app:data_sources} describes the data sources and mixture construction, and Appendix~\ref{app:prompts} lists the target categories.

We evaluate \texttt{Qwen2-Audio}, \texttt{Kimi-Audio}, and \texttt{Step-Audio-2-mini}~\citep{chu2024qwen2,ding2025kimi,wu2025step}.
We construct long-term memory in each ALLM's input embedding space without training the model, using $N=20$ separate clean reference clips per category for the aggregate accuracy comparisons.
Detailed results across memory sizes are provided in Appendix~\ref{app:detailed_results}.

\begin{figure}[t]
\centering

\includegraphics[
    width=\columnwidth
]{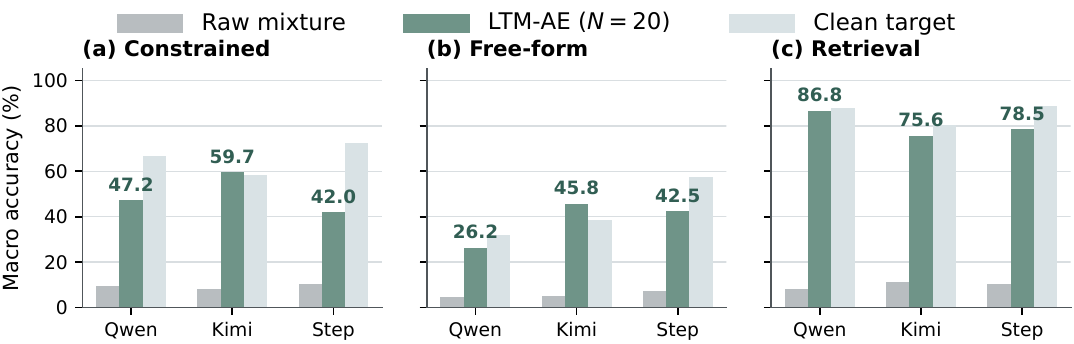}

\caption{\textbf{Diagnostic readouts under acoustic interference.}
Macro-average accuracy of \texttt{Qwen2-Audio}, \texttt{Kimi-Audio}, and \texttt{Step-Audio-2-mini}
on clean audio, raw mixtures, and LTM-AE-enhanced mixtures, evaluated through three readouts: constrained classification, free-form classification, and prompt-free retrieval.}

\label{fig:final20_main}
\vspace{-9pt}
\end{figure}

\Needspace{24\baselineskip}
\subsection{Diagnostic Readouts of Selective Target Perception}
\label{sec:final20_results}

\begin{wraptable}{r}{0.52\textwidth}
\vspace{-\intextsep}
\setlength{\abovecaptionskip}{0pt}
\centering
\caption{Macro-average accuracy (\%) over the twenty audio categories.
C denotes constrained classification, F denotes free-form classification,
and R denotes prompt-free retrieval.}
\label{tab:final20_main}
\small
\setlength{\tabcolsep}{3pt}
\begin{tabular}{@{}llrrr@{}}
\toprule
Model & Readout & Clean & Mixed & LTM-AE \\
\midrule
Qwen2-Audio & C & 66.90 & 9.60  & 47.20 \\
            & F & 32.05 & 4.80  & 26.25 \\
            & R & 87.90 & 8.25  & 86.75 \\
\midrule
Kimi-Audio & C & 58.20 & 8.30  & 59.70 \\
           & F & 38.45 & 4.90  & 45.80 \\
           & R & 80.40 & 11.25 & 75.60 \\
\midrule
Step-Audio-2-mini & C & 72.50 & 10.25 & 42.05 \\
                  & F & 57.30 & 7.15  & 42.45 \\
                  & R & 88.95 & 10.45 & 78.50 \\
\bottomrule
\end{tabular}
\vspace{-\baselineskip}
\end{wraptable}

To assess responses to the specified target under long-term memory guidance, we compare clean recordings, raw mixtures, and LTM-AE-enhanced mixtures through three diagnostic readouts: constrained classification, free-form classification, and prompt-free retrieval.
For LTM-AE, all three readouts use target-conditioned enhancement, with the specified category selecting the long-term memory.
The classification prompts do not identify the selected target, and retrieval receives no separate target cue.
The results are shown in Figure~\ref{fig:final20_main} and Table~\ref{tab:final20_main}.
Clean recordings establish performance when the target is unobscured, while raw mixtures are evaluated without category guidance.
Constrained classification measures how often the model reports the specified category from a fixed answer space.
Free-form classification measures the same category response without a candidate list.
Prompt-free retrieval is a representation-level diagnostic of alignment with the specified target category against a separate clean reference gallery.
Together, these readouts measure how category-conditioned enhancement changes responses to the specified target in audio representations and language outputs.

\paragraph{Language-output results.}
LTM-AE improves constrained classification across all three ALLMs.
For \texttt{Qwen2-Audio}, accuracy increases from 9.60\% on raw mixtures to 47.20\% with LTM-AE.
\texttt{Kimi-Audio} improves from 8.30\% to 59.70\%, while \texttt{Step-Audio-2-mini} improves from 10.25\% to 42.05\%.
Removing the candidate list retains the benefit across all three models.
\texttt{Qwen2-Audio}'s free-form classification accuracy rises from 4.80\% to 26.25\%.
\texttt{Kimi-Audio} reaches 45.80\% from a raw-mixture accuracy of 4.90\%, and \texttt{Step-Audio-2-mini} rises from 7.15\% to 42.45\%.
The gains in both output formats occur with unchanged prompts and without training the ALLM.
On 100 inputs without piano, \texttt{Qwen2-Audio} produces no piano mentions in its free-form responses either before or after enhancement with piano long-term memory (Appendix~\ref{app:target_absence}).

\paragraph{Representation-level retrieval results.}
The improvement is also present in the representation-level retrieval diagnostic.
For \texttt{Qwen2-Audio}, accuracy rises from 8.25\% on raw mixtures to 86.75\%, approaching its 87.90\% accuracy on clean recordings.
\texttt{Kimi-Audio} and \texttt{Step-Audio-2-mini} also improve under this protocol.
Since retrieval bypasses language generation, these gains show stronger alignment with clean recordings of the specified category before the language backbone decodes the audio.

\paragraph{Takeaway.}
With the target category specified, LTM-AE strengthens responses to this target in four-source mixtures across all three ALLMs without training them.
The gains appear in both audio representations and language outputs.

\begin{figure}[t]
\centering
\includegraphics[width=\columnwidth]{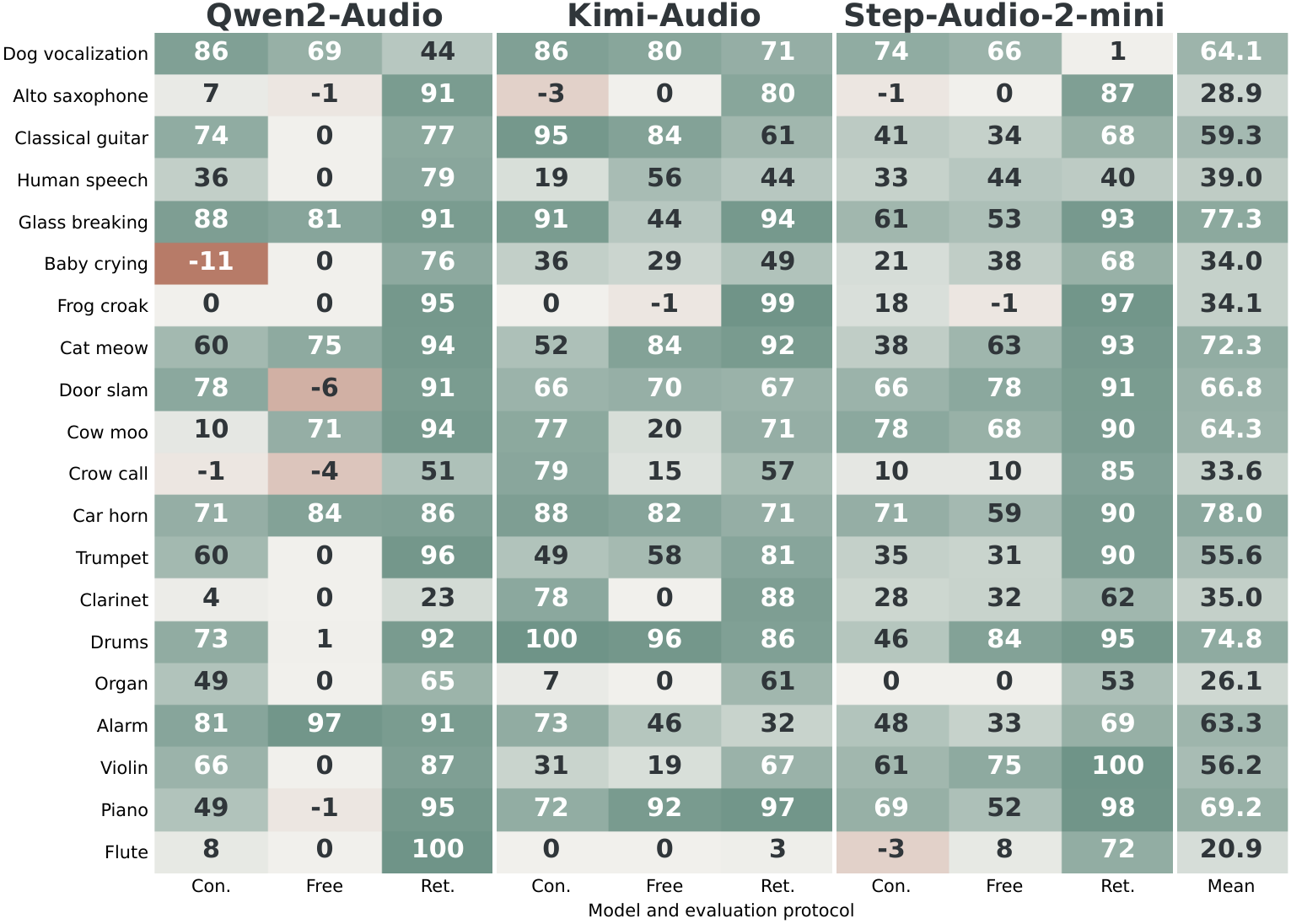}
\caption{\textbf{MultiPerception class-wise accuracy gains.}
Each cell reports the best accuracy across the evaluated memory sizes minus the raw-mixture accuracy, in percentage points (Appendix~\ref{app:detailed_results}).
Columns correspond to \texttt{Qwen2-Audio}, \texttt{Kimi-Audio}, and \texttt{Step-Audio-2-mini} under constrained classification, free-form classification, and prompt-free retrieval.
% The final column averages the nine gains for each category.
}
\label{fig:final20_classwise_gain_main}
\vspace{-13.5pt}
\end{figure}

\FloatBarrier

% Reserve space for the heading and the complete wrapped table.
\Needspace{24\baselineskip}
\subsection{Analysis of Enhancement Gains across Readouts}
\label{sec:enhancement_gain_analysis}

\begin{wraptable}{r}{0.53\textwidth}
\vspace{-\intextsep}
\setlength{\abovecaptionskip}{0pt}
\centering
\caption{Accuracy gains in percentage points relative to the raw-mixture condition. The mean language-output gain averages constrained and free-form classification.}
\label{tab:gain_analysis}
\small
\setlength{\tabcolsep}{4pt}
\begin{tabular}{@{}lrrr@{}}
\toprule
Readout & \shortstack{Qwen2-\\Audio} & \shortstack{Kimi-\\Audio} & \shortstack{Step-Audio-\\2-mini} \\
\midrule
\begin{tabular}[c]{@{}l@{}}Constrained\\classification\end{tabular} & +37.60 & +51.40 & +31.80 \\
\begin{tabular}[c]{@{}l@{}}Free-form\\classification\end{tabular}   & +21.45 & +40.90 & +35.30 \\
\midrule
\begin{tabular}[c]{@{}l@{}}Mean language-\\output gain\end{tabular}  & +29.53 & +46.15 & +33.55 \\
\begin{tabular}[c]{@{}l@{}}Prompt-free\\retrieval\end{tabular}      & +78.50 & +64.35 & +68.05 \\
\bottomrule
\end{tabular}
\vspace{-\baselineskip}
\end{wraptable}

To examine how broadly LTM-AE improves the diagnostic readouts, we compare gains across model interfaces and sound categories in Table~\ref{tab:gain_analysis} and Figure~\ref{fig:final20_classwise_gain_main}.
Gains are measured in percentage points relative to raw mixtures.

\paragraph{Cross-model comparison.}
LTM-AE improves all nine combinations of model and readout.
\texttt{Kimi-Audio} obtains the largest mean gain across the two language-output protocols at 46.15 percentage points.
\texttt{Qwen2-Audio} and \texttt{Step-Audio-2-mini} gain 29.53 and 33.55 points, respectively.
Even the smallest language-output gain is 21.45 points, achieved by \texttt{Qwen2-Audio} in free-form classification.
The improvement therefore persists across different ALLM interfaces, with each model using long-term memory constructed in its own input embedding space.
LTM-AE turns clean auditory experience into stronger responses to the specified target across these models without training them.

\paragraph{From representation to language output.}
Retrieval gains are larger than language-output gains for every ALLM.
\texttt{Qwen2-Audio} gains 78.50 points in retrieval, while \texttt{Step-Audio-2-mini} and \texttt{Kimi-Audio} gain 68.05 and 64.35 points.
\texttt{Qwen2-Audio} thus benefits most under direct retrieval, whereas \texttt{Kimi-Audio} benefits most in language output.
This difference shows that the gains from refining audio embeddings do not translate uniformly across readouts.
Nevertheless, the positive gains in both language-output protocols establish that the enhanced representations remain useful after decoding.

\paragraph{Perception beyond speech.}
Figure~\ref{fig:final20_classwise_gain_main} resolves these gains by target category, using the best observed result across the evaluated memory sizes for each model and readout.
The retrieval columns are positive for every category in all three ALLMs, and this coverage also holds under the setup used for the aggregate comparisons (Appendix~\ref{app:detailed_results}).
Car horn and glass breaking have mean gains of 78.0 and 77.3 points across the nine combinations, with improvement in every combination.
For car horn, the free-form gains reach 84 points for \texttt{Qwen2-Audio}, 82 for \texttt{Kimi-Audio}, and 59 for \texttt{Step-Audio-2-mini}.
These gains show that the models can express the benefit of long-term memory guidance in their own descriptions of non-speech targets.
The effect also extends to musical instruments and animal sounds.
For drums, \texttt{Kimi-Audio} reaches 100\% in both classification protocols, compared with 0\% in constrained classification and 4\% in free-form classification on raw mixtures.
Cat-meow retrieval reaches 100\% for all three ALLMs.
LTM-AE therefore strengthens responses to the specified target across sound families when multiple competing sources obscure the target sound.

\paragraph{Category alignment before decoding.}
The heatmap also reveals improvements that language outputs alone would miss.
Frog-croak retrieval gains reach 95 points for \texttt{Qwen2-Audio}, 99 for \texttt{Kimi-Audio}, and 97 for \texttt{Step-Audio-2-mini}, although free-form classification does not improve for this category.
LTM-AE thus strengthens alignment with the specified category in the audio embeddings even when the ALLM still fails to name the sound in an open-ended response.

\paragraph{Takeaway.}
Across twenty sound categories, LTM-AE improves retrieval for every ALLM and delivers language-output gains for diverse non-speech targets.

\FloatBarrier

\subsection{Extending Long-Term Memory Guidance to Speech Transcription}
\label{sec:speech_transcription}

To assess whether a gated extension of LTM-AE helps recover the selected target's content, we conduct an experiment on speech transcription.
The speech category specifies the listening target, while the words spoken in the current recording remain unknown and must be recovered from the mixture.
We examine whether allowing the strength of long-term memory guidance to vary across audio tokens helps retain this information.
\begin{figure}[!t]
\centering
\includegraphics[
    width=\columnwidth
]{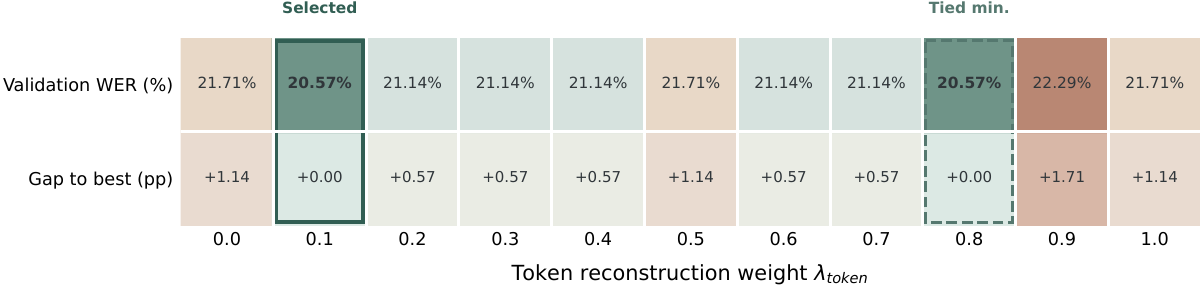}

\caption{\textbf{Validation-based selection of the token-level gate.}
WER on the 20-example selection split for
$\lambda_{\mathrm{tok}}\in\{0.0,0.1,\ldots,1.0\}$.
}

\label{fig:token_gate_selection}
\end{figure}

This experiment uses \texttt{Qwen2-Audio} and English LibriSpeech mixtures, with a separate speech long-term memory whose settings are detailed in Appendix~\ref{app:speech_results}.
The speech long-term memory and all ALLM parameters remain fixed.
For this extension, we train the token-level gate described in Appendix~\ref{sec:memory_gate} to scale the correction from long-term memory at each temporal position.
A gate value near zero retains more of the original token, while a value near one applies more of the correction.

We divide 100 development mixtures into 80 training examples and 20 selection examples.
The gate is trained with a transcription loss and a token reconstruction loss that compares its output embeddings with those of clean target recordings.
For each
$\lambda_{\mathrm{tok}}\in\{0.0,0.1,\ldots,1.0\}$, we train the gate for four
epochs with a learning rate of $10^{-3}$ and
$\lambda_{\mathrm{text}}=1$, and compute word error rate (WER) on the selection split.
Figure~\ref{fig:token_gate_selection} shows the selection WER for each token reconstruction weight, with the second row giving its gap to the lowest WER in percentage points.
Both $\lambda_{\mathrm{tok}}=0.1$ and $0.8$ reach this minimum of 20.57\%.
We select $0.1$ according to the predefined rule that chooses the smaller weight when selection performance is tied.

We then retrain the selected configuration on all 100 development examples
and evaluate it once on 100 held-out test mixtures.
WER is a standard metric for evaluating automatic speech recognition~\citep{morris2004and,panayotov2015librispeech,radford2023robust}.
We compute it as
\begin{equation}
\operatorname{WER}
=
\frac{S+D+I}{R},
\label{eq:wer}
\end{equation}
where $S$, $D$, and $I$ denote substitutions, deletions, and insertions,
and $R$ is the number of words in the normalized reference transcript.

\Needspace{16\baselineskip}
\begin{wraptable}{r}{0.50\textwidth}
\vspace{-\intextsep}
\setlength{\abovecaptionskip}{0pt}
\centering
\caption{\textbf{Speech transcription results.} WER on the matched held-out test set. Lower is better.}
\label{tab:speech_wer_formal}
\small
\setlength{\tabcolsep}{3pt}
\renewcommand{\arraystretch}{1.05}
\begin{tabular*}{\linewidth}{@{\extracolsep{\fill}}lrr@{}}
\toprule
Condition & WER (\%) & $\Delta$ raw \\
\midrule
Raw mixture & 23.07 & 0.00 \\
Ungated LTM-AE & 33.75 & +10.68 \\
Selected gate & \textbf{14.77} & \textbf{$-8.30$} \\
\bottomrule
\end{tabular*}
\vspace{-\baselineskip}
\end{wraptable}

On the held-out test set, the selected gate reduces WER from 23.07\% on raw mixtures to 14.77\% across 880 reference words (Table~\ref{tab:speech_wer_formal}).
The gated LTM-AE configuration thus improves recovery of the words spoken in the current recording.

LTM-AE without the learned gate yields 33.75\% WER, exceeding the raw-mixture baseline.
In this setting, using one interpolation weight for all tokens is less effective than allowing the gate to adjust it at each position.

\paragraph{Takeaway.}
LTM-AE with an additional learned gate on \texttt{Qwen2-Audio} improves recovery of the selected target's spoken content, which is not specified by the speech category.
Further work could examine how long-term memory guidance extends to other audio tasks and ALLMs.
\FloatBarrier

\section{Conclusions}

We introduced LTM-AE to support selective perception in complex audio mixtures through guidance inspired by human long-term memory.
By retaining category structure from clean audio hidden states, LTM-AE turns prior auditory experience into reusable guidance for refining incoming audio embeddings without training the ALLM.
Experiments across three ALLMs and twenty sound categories demonstrate improvements in language outputs and closer alignment of audio representations with the specified target category.
These gains extend to diverse non-speech sounds under interference from multiple sources.
An extension with a learned token-level gate also improves speech transcription, motivating further work on adapting long-term memory guidance to audio tasks that require the specific content of a recording.

\section*{AI Use Statement}

Generative AI tools were used to assist with literature retrieval and reference verification, review manuscript formatting and consistency, provide feedback on research methodology and experimental reporting, and draft and revise manuscript text. All AI-assisted outputs and suggestions were reviewed by the authors. The authors take full responsibility for the final content of this work, including all text, data, results, and claims.

\section*{Reproducibility Statement}

To support reproducibility, we release our implementation and evaluation code in the Supplementary Material. Details of data preparation, memory construction, experimental settings, and evaluation protocols are provided in the appendix.

\bibliographystyle{iclr2027_conference}
\bibliography{custom}

\clearpage

\appendix

\section{Detailed Twenty-Class Results}
\label{app:detailed_results}

This appendix reports the complete class-wise results of \texttt{Qwen2-Audio}, \texttt{Kimi-Audio}, and \texttt{Step-Audio-2-mini} under constrained classification, free-form classification, and prompt-free retrieval.
The PCA columns report LTM-AE with long-term memory constructed from each evaluated reference count, with the retained rank and interpolation weight calibrated as described in Section~\ref{sec:memory_calibration}.
Each row measures responses to the specified target category, and the final row averages the twenty category accuracies.
Reading across the three protocols connects changes in the audio representations with the categories expressed by the language backbone.
The complete results also show how these responses vary with the number of clean reference clips used to construct long-term memory.
\subsection{\texttt{Qwen2-Audio}}

For \texttt{Qwen2-Audio}, LTM-AE improves the mean accuracy of all three readouts at every evaluated reference count (Tables~\ref{tab:qwen_constrained_classwise}--\ref{tab:qwen_retrieval_classwise}).
These gains include environmental targets that are rarely reported from raw mixtures.
For example, constrained accuracy for glass breaking rises from 12\% to 100\% at all three reference counts, while its free-form accuracy rises from 4\% to 50--85\%.
Free-form accuracy for car horn reaches 100\% from 16\% in the raw condition across all three reference counts.
The retrieval diagnostic provides a corresponding view of category alignment in the enhanced embeddings.
For piano, drums, and glass breaking, retrieval reaches 100\% at every reference count, compared with raw accuracies of 5\%, 8\%, and 9\%, respectively.
These results illustrate how the same enhancement procedure strengthens responses to selected targets across musical and environmental sounds, both before decoding and in generated descriptions.

\begin{table}[h]
\centering
\caption{Class-wise constrained-classification accuracy (\%) of Qwen2-Audio.}
\label{tab:qwen_constrained_classwise}
\scriptsize
\setlength{\tabcolsep}{3pt}
\resizebox{\columnwidth}{!}{%
\begin{tabular}{lccccc}
\hline
Class & Clean & Mixed & PCA ($N=20$) & PCA ($N=50$) & PCA ($N=100$) \\
\hline
Piano             & 85 & 0  & 38  & 49  & 44  \\
Dog vocalization  & 86 & 14 & 100 & 100 & 99  \\
Human speech      & 94 & 64 & 100 & 100 & 100 \\
Flute             & 43 & 1  & 5   & 9   & 5   \\
Organ             & 61 & 1  & 38  & 50  & 35  \\
Classical guitar  & 87 & 3  & 63  & 77  & 64  \\
Violin            & 92 & 8  & 67  & 73  & 74  \\
Trumpet           & 64 & 3  & 63  & 11  & 17  \\
Clarinet          & 55 & 0  & 1   & 1   & 4   \\
Alto saxophone    & 24 & 3  & 0   & 10  & 0   \\
Drums             & 87 & 1  & 63  & 55  & 74  \\
Cat meow          & 100 & 40 & 100 & 100 & 100 \\
Cow moo           & 84 & 7  & 15  & 13  & 17  \\
Frog croak        & 0  & 0  & 0   & 0   & 0   \\
Crow call         & 29 & 1  & 0   & 0   & 0   \\
Glass breaking    & 88 & 12 & 100 & 100 & 100 \\
Car horn          & 77 & 7  & 27  & 61  & 78  \\
Baby crying       & 39 & 14 & 0   & 0   & 3   \\
Door slam         & 91 & 10 & 88  & 86  & 86  \\
Alarm             & 52 & 3  & 76  & 84  & 84  \\
\hline
Average           & 66.90 & 9.60 & 47.20 & 48.95 & 49.20 \\
\hline
\end{tabular}
}
\end{table}
\begin{table}[H]
\centering
\caption{Class-wise free-form classification accuracy (\%) of Qwen2-Audio.}
\label{tab:qwen_freeform_classwise}
\scriptsize
\setlength{\tabcolsep}{3pt}
\resizebox{\columnwidth}{!}{%
\begin{tabular}{lccccc}
\hline
Class & Clean & Mixed & PCA ($N=20$) & PCA ($N=50$) & PCA ($N=100$) \\
\hline
Piano             & 0  & 1  & 0   & 0   & 0   \\
Dog vocalization  & 83 & 31 & 100 & 100 & 100 \\
Human speech      & 3  & 0  & 0   & 0   & 0   \\
Flute             & 2  & 0  & 0   & 0   & 0   \\
Organ             & 38 & 0  & 0   & 0   & 0   \\
Classical guitar  & 5  & 1  & 1   & 0   & 0   \\
Violin            & 3  & 0  & 0   & 0   & 0   \\
Trumpet           & 15 & 0  & 0   & 0   & 0   \\
Clarinet          & 0  & 0  & 0   & 0   & 0   \\
Alto saxophone    & 2  & 1  & 0   & 0   & 0   \\
Drums             & 37 & 0  & 0   & 1   & 1   \\
Cat meow          & 89 & 20 & 95  & 91  & 89  \\
Cow moo           & 79 & 9  & 79  & 52  & 80  \\
Frog croak        & 0  & 0  & 0   & 0   & 0   \\
Crow call         & 38 & 4  & 0   & 0   & 0   \\
Glass breaking    & 72 & 4  & 50  & 82  & 85  \\
Car horn          & 83 & 16 & 100 & 100 & 100 \\
Baby crying       & 0  & 0  & 0   & 0   & 0   \\
Door slam         & 48 & 6  & 0   & 0   & 0   \\
Alarm             & 44 & 3  & 100 & 98  & 97  \\
\hline
Average           & 32.05 & 4.80 & 26.25 & 26.20 & 27.60 \\
\hline
\end{tabular}
}
\end{table}

\begin{table}[H]
\centering
\caption{Class-wise prompt-free retrieval accuracy (\%) of Qwen2-Audio.}
\label{tab:qwen_retrieval_classwise}
\scriptsize
\setlength{\tabcolsep}{3pt}
\resizebox{\columnwidth}{!}{%
\begin{tabular}{lccccc}
\hline
Class & Clean & Mixed & PCA ($N=20$) & PCA ($N=50$) & PCA ($N=100$) \\
\hline
Piano             & 99  & 5  & 100 & 100 & 100 \\
Dog vocalization  & 75  & 0  & 44  & 20  & 6   \\
Human speech      & 100 & 21 & 100 & 100 & 100 \\
Flute             & 76  & 0  & 100 & 100 & 100 \\
Organ             & 87  & 0  & 61  & 65  & 64  \\
Classical guitar  & 91  & 23 & 100 & 100 & 100 \\
Violin            & 95  & 0  & 82  & 83  & 87  \\
Trumpet           & 77  & 4  & 100 & 90  & 90  \\
Clarinet          & 88  & 3  & 13  & 16  & 26  \\
Alto saxophone    & 79  & 9  & 98  & 100 & 98  \\
Drums             & 93  & 8  & 100 & 100 & 100 \\
Cat meow          & 96  & 6  & 100 & 100 & 100 \\
Cow moo           & 97  & 6  & 100 & 100 & 100 \\
Frog croak        & 100 & 5  & 100 & 100 & 100 \\
Crow call         & 78  & 10 & 61  & 61  & 60  \\
Glass breaking    & 95  & 9  & 100 & 100 & 100 \\
Car horn          & 83  & 14 & 100 & 100 & 100 \\
Baby crying       & 86  & 24 & 100 & 100 & 100 \\
Door slam         & 92  & 9  & 100 & 100 & 100 \\
Alarm             & 71  & 9  & 76  & 100 & 91  \\
\hline
Average           & 87.90 & 8.25 & 86.75 & 86.75 & 86.10 \\
\hline
\end{tabular}
}
\end{table}
\subsection{\texttt{Kimi-Audio}}

\texttt{Kimi-Audio} shows gains in both language-output protocols across the evaluated reference counts (Tables~\ref{tab:kimi_constrained_classwise} and~\ref{tab:kimi_freeform_classwise}).
Constrained accuracy rises from 8.30\% on raw mixtures to 57.60--59.70\%, while free-form accuracy rises from 4.90\% to 43.30--45.80\%.
The category results show that these gains extend to descriptions generated without a candidate list.
Free-form accuracy for classical guitar improves from 13\% to 95--97\%, and piano improves from 0\% to 85--92\%.
For cat meows, free-form accuracy reaches 100\% at all three reference counts from a raw accuracy of 16\%.
The same category reaches 100\% in retrieval, compared with 8\% on raw mixtures (Table~\ref{tab:kimi_retrieval_classwise}).
These cases connect stronger alignment in the audio representation with more frequent descriptions of the selected target, using category long-term memory constructed in \texttt{Kimi-Audio}'s own input embedding space.
\begin{table}[H]
\centering
\caption{Class-wise constrained-classification accuracy (\%) of Kimi-Audio.}
\label{tab:kimi_constrained_classwise}
\scriptsize
\setlength{\tabcolsep}{3pt}
\resizebox{\columnwidth}{!}{%
\begin{tabular}{lccccc}
\hline
Class & Clean & Mixed & PCA ($N=20$) & PCA ($N=50$) & PCA ($N=100$) \\
\hline
Piano             & 100 & 28 & 100 & 100 & 100 \\
Dog vocalization  & 90  & 14 & 100 & 91  & 84  \\
Human speech      & 84  & 10 & 29  & 29  & 26  \\
Flute             & 2   & 0  & 0   & 0   & 0   \\
Organ             & 13  & 4  & 5   & 10  & 11  \\
Classical guitar  & 91  & 4  & 96  & 98  & 99  \\
Violin            & 12  & 3  & 34  & 10  & 26  \\
Trumpet           & 45  & 0  & 49  & 35  & 20  \\
Clarinet          & 79  & 22 & 99  & 98  & 100 \\
Alto saxophone    & 1   & 3  & 0   & 0   & 0   \\
Drums             & 75  & 0  & 75  & 100 & 100 \\
Cat meow          & 66  & 16 & 60  & 65  & 68  \\
Cow moo           & 91  & 12 & 76  & 88  & 89  \\
Frog croak        & 0   & 0  & 0   & 0   & 0   \\
Crow call         & 45  & 5  & 84  & 77  & 75  \\
Glass breaking    & 95  & 9  & 100 & 100 & 99  \\
Car horn          & 72  & 6  & 88  & 94  & 92  \\
Baby crying       & 36  & 7  & 37  & 39  & 43  \\
Door slam         & 92  & 4  & 70  & 46  & 41  \\
Alarm             & 75  & 19 & 92  & 85  & 79  \\
\hline
Average           & 58.20 & 8.30 & 59.70 & 58.25 & 57.60 \\
\hline
\end{tabular}
}
\end{table}
\begin{table}[H]
\centering
\caption{Class-wise free-form classification accuracy (\%) of Kimi-Audio.}
\label{tab:kimi_freeform_classwise}
\scriptsize
\setlength{\tabcolsep}{3pt}
\resizebox{\columnwidth}{!}{%
\begin{tabular}{lccccc}
\hline
Class & Clean & Mixed & PCA ($N=20$) & PCA ($N=50$) & PCA ($N=100$) \\
\hline
Piano             & 46  & 0  & 90  & 92  & 85 \\
Dog vocalization  & 76  & 17 & 97  & 73  & 62 \\
Human speech      & 100 & 12 & 68  & 65  & 68 \\
Flute             & 1   & 0  & 0   & 0   & 0  \\
Organ             & 1   & 0  & 0   & 0   & 0  \\
Classical guitar  & 62  & 13 & 97  & 95  & 96 \\
Violin            & 18  & 1  & 17  & 5   & 20 \\
Trumpet           & 15  & 0  & 58  & 48  & 55 \\
Clarinet          & 0   & 0  & 0   & 0   & 0  \\
Alto saxophone    & 0   & 0  & 0   & 0   & 0  \\
Drums             & 78  & 4  & 82  & 100 & 99 \\
Cat meow          & 91  & 16 & 100 & 100 & 100 \\
Cow moo           & 12  & 0  & 17  & 19  & 20 \\
Frog croak        & 0   & 1  & 0   & 0   & 0  \\
Crow call         & 14  & 0  & 3   & 15  & 12 \\
Glass breaking    & 55  & 0  & 44  & 34  & 29 \\
Car horn          & 43  & 6  & 70  & 88  & 80 \\
Baby crying       & 40  & 18 & 47  & 43  & 44 \\
Door slam         & 82  & 7  & 77  & 74  & 60 \\
Alarm             & 35  & 3  & 49  & 34  & 36 \\
\hline
Average           & 38.45 & 4.90 & 45.80 & 44.25 & 43.30 \\
\hline
\end{tabular}
}
\end{table}
\begin{table}[H]
\centering
\caption{Class-wise prompt-free retrieval accuracy (\%) of Kimi-Audio.}
\label{tab:kimi_retrieval_classwise}
\scriptsize
\setlength{\tabcolsep}{3pt}
\resizebox{\columnwidth}{!}{%
\begin{tabular}{lccccc}
\hline
Class & Clean & Mixed & PCA ($N=20$) & PCA ($N=50$) & PCA ($N=100$) \\
\hline
Piano             & 96  & 3  & 100 & 100 & 99 \\
Dog vocalization  & 63  & 0  & 71  & 26  & 32 \\
Human speech      & 100 & 53 & 97  & 89  & 91 \\
Flute             & 35  & 0  & 3   & 2   & 2  \\
Organ             & 81  & 1  & 43  & 42  & 62 \\
Classical guitar  & 88  & 25 & 86  & 83  & 78 \\
Violin            & 70  & 12 & 71  & 79  & 79 \\
Trumpet           & 84  & 8  & 89  & 77  & 84 \\
Clarinet          & 83  & 0  & 79  & 88  & 80 \\
Alto saxophone    & 85  & 7  & 79  & 82  & 87 \\
Drums             & 94  & 2  & 88  & 85  & 85 \\
Cat meow          & 93  & 8  & 100 & 100 & 100 \\
Cow moo           & 98  & 27 & 98  & 98  & 98 \\
Frog croak        & 100 & 1  & 100 & 100 & 100 \\
Crow call         & 60  & 3  & 33  & 60  & 53 \\
Glass breaking    & 97  & 6  & 100 & 100 & 94 \\
Car horn          & 58  & 7  & 69  & 64  & 78 \\
Baby crying       & 88  & 42 & 91  & 87  & 84 \\
Door slam         & 84  & 8  & 71  & 74  & 75 \\
Alarm             & 51  & 12 & 44  & 16  & 25 \\
\hline
Average           & 80.40 & 11.25 & 75.60 & 72.60 & 74.30 \\
\hline
\end{tabular}
}
\end{table}
\subsection{\texttt{Step-Audio-2-mini}}

For \texttt{Step-Audio-2-mini}, mean constrained accuracy increases from 10.25\% on raw mixtures to 42.05--45.80\%, and free-form accuracy increases from 7.15\% to 42.45--44.65\% (Tables~\ref{tab:stepfun_constrained_classwise} and~\ref{tab:stepfun_freeform_classwise}).
Violin provides an example of improvement under both output formats.
Its constrained accuracy rises from 6\% to 57--67\%, and its free-form accuracy rises from 4\% to 62--79\%.
The language-output gains also cover animal and environmental sounds.
Free-form accuracy for cow moo increases from 3\% to 65--71\%, while door slam increases from 14\% to 76--92\%.
Retrieval accuracy improves from 10.45\% overall to 78.50--83.65\% (Table~\ref{tab:stepfun_retrieval_classwise}), with cat meow, cow moo, and car horn reaching 100\% at every reference count.
Together with the other two ALLMs, these results demonstrate that LTM-AE improves the diagnostic responses across different audio interfaces without training the underlying models.

\begin{table}[H]
\centering
\caption{Class-wise constrained-classification accuracy (\%) of Step-Audio-2-mini.}
\label{tab:stepfun_constrained_classwise}
\scriptsize
\setlength{\tabcolsep}{3pt}
\resizebox{\columnwidth}{!}{%
\begin{tabular}{lccccc}
\hline
Class & Clean & Mixed & PCA ($N=20$) & PCA ($N=50$) & PCA ($N=100$) \\
\hline
Piano             & 100 & 0  & 29 & 58 & 69 \\
Dog vocalization  & 90  & 23 & 96 & 97 & 92 \\
Human speech      & 98  & 65 & 98 & 97 & 95 \\
Flute             & 32  & 4  & 0  & 0  & 1  \\
Organ             & 30  & 2  & 1  & 1  & 2  \\
Classical guitar  & 92  & 7  & 48 & 46 & 46 \\
Violin            & 96  & 6  & 57 & 64 & 67 \\
Trumpet           & 68  & 3  & 26 & 34 & 38 \\
Clarinet          & 90  & 5  & 15 & 30 & 33 \\
Alto saxophone    & 14  & 1  & 0  & 0  & 0  \\
Drums             & 90  & 2  & 48 & 31 & 43 \\
Cat meow          & 96  & 22 & 56 & 58 & 60 \\
Cow moo           & 91  & 8  & 80 & 80 & 86 \\
Frog croak        & 48  & 2  & 19 & 20 & 20 \\
Crow call         & 46  & 4  & 14 & 10 & 13 \\
Glass breaking    & 90  & 12 & 57 & 73 & 54 \\
Car horn          & 75  & 7  & 71 & 78 & 51 \\
Baby crying       & 41  & 16 & 27 & 28 & 37 \\
Door slam         & 92  & 10 & 45 & 76 & 48 \\
Alarm             & 71  & 6  & 54 & 35 & 41 \\
\hline
Average           & 72.50 & 10.25 & 42.05 & 45.80 & 44.80 \\
\hline
\end{tabular}
}
\end{table}
\begin{table}[H]
\centering
\caption{Class-wise free-form classification accuracy (\%) of Step-Audio-2-mini.}
\label{tab:stepfun_freeform_classwise}
\scriptsize
\setlength{\tabcolsep}{3pt}
\resizebox{\columnwidth}{!}{%
\begin{tabular}{lccccc}
\hline
Class & Clean & Mixed & PCA ($N=20$) & PCA ($N=50$) & PCA ($N=100$) \\
\hline
Piano             & 78 & 2  & 54 & 49 & 29 \\
Dog vocalization  & 80 & 26 & 87 & 92 & 89 \\
Human speech      & 80 & 19 & 55 & 63 & 62 \\
Flute             & 22 & 0  & 1  & 1  & 8  \\
Organ             & 2  & 0  & 0  & 0  & 0  \\
Classical guitar  & 66 & 10 & 44 & 37 & 30 \\
Violin            & 81 & 4  & 62 & 75 & 79 \\
Trumpet           & 52 & 1  & 13 & 20 & 32 \\
Clarinet          & 71 & 3  & 25 & 28 & 35 \\
Alto saxophone    & 31 & 6  & 2  & 2  & 6  \\
Drums             & 65 & 4  & 88 & 74 & 79 \\
Cat meow          & 98 & 24 & 84 & 86 & 87 \\
Cow moo           & 79 & 3  & 65 & 66 & 71 \\
Frog croak        & 0  & 1  & 0  & 0  & 0  \\
Crow call         & 29 & 0  & 10 & 8  & 10 \\
Glass breaking    & 80 & 1  & 40 & 54 & 33 \\
Car horn          & 80 & 7  & 66 & 58 & 65 \\
Baby crying       & 29 & 13 & 39 & 34 & 51 \\
Door slam         & 83 & 14 & 76 & 92 & 90 \\
Alarm             & 40 & 5  & 38 & 20 & 37 \\
\hline
Average           & 57.30 & 7.15 & 42.45 & 42.95 & 44.65 \\
\hline
\end{tabular}
}
\end{table}
\begin{table}[H]
\centering
\caption{Class-wise prompt-free retrieval accuracy (\%) of Step-Audio-2-mini.}
\label{tab:stepfun_retrieval_classwise}
\scriptsize
\setlength{\tabcolsep}{3pt}
\resizebox{\columnwidth}{!}{%
\begin{tabular}{lccccc}
\hline
Class & Clean & Mixed & PCA ($N=20$) & PCA ($N=50$) & PCA ($N=100$) \\
\hline
Piano             & 99  & 2  & 100 & 100 & 100 \\
Dog vocalization  & 74  & 0  & 1   & 0   & 0   \\
Human speech      & 100 & 60 & 100 & 100 & 100 \\
Flute             & 72  & 0  & 50  & 51  & 72  \\
Organ             & 87  & 0  & 32  & 41  & 53  \\
Classical guitar  & 89  & 32 & 61  & 100 & 100 \\
Violin            & 94  & 0  & 100 & 91  & 83  \\
Trumpet           & 81  & 1  & 91  & 85  & 91  \\
Clarinet          & 90  & 2  & 26  & 37  & 64  \\
Alto saxophone    & 82  & 13 & 99  & 100 & 100 \\
Drums             & 93  & 5  & 100 & 100 & 100 \\
Cat meow          & 99  & 7  & 100 & 100 & 100 \\
Cow moo           & 98  & 10 & 100 & 100 & 100 \\
Frog croak        & 100 & 3  & 100 & 100 & 100 \\
Crow call         & 81  & 9  & 94  & 94  & 89  \\
Glass breaking    & 98  & 7  & 56  & 100 & 60  \\
Car horn          & 86  & 10 & 100 & 100 & 100 \\
Baby crying       & 88  & 31 & 99  & 97  & 96  \\
Door slam         & 91  & 9  & 100 & 100 & 97  \\
Alarm             & 77  & 8  & 61  & 77  & 45  \\
\hline
Average           & 88.95 & 10.45 & 78.50 & 83.65 & 82.50 \\
\hline
\end{tabular}
}
\end{table}
\clearpage
\section{Prompts and Evaluation Details}
\label{app:prompts}

The twenty-class experiments use the same prompt templates and candidate labels for all three ALLMs. These prompts do not identify which category is selected for enhancement. LTM-AE uses the listening cue to select the long-term memory and its calibrated settings, then passes the enhanced audio embeddings to the readout without a separate target cue. We provide the prompts here, followed by the retrieval procedure and the conditions used for comparison.

\subsection{Constrained-Classification Prompt}

Constrained classification measures the model's response to the specified target within a fixed answer space. The same candidate list is used for clean recordings, raw mixtures, and enhanced mixtures. The following prompt requires the model to select exactly one category from the predefined label set:

\begin{quote}
% \small
\texttt{Classify the main sound in this audio. Choose exactly one option.}

\texttt{Reply with only one capital letter from A to T.}
\end{quote}

\begin{table}[H]
\centering
\caption{Candidate labels used for constrained classification.}
\label{tab:candidate_labels}
\small
\setlength{\tabcolsep}{5pt}
\begin{tabular}{clcl}
\hline
Label & Sound category & Label & Sound category \\
\hline
A & Piano music            & K & Concert drums \\
B & Dog vocalization       & L & Cat meow \\
C & Human speech           & M & Cow moo \\
D & Flute music            & N & Frog croak \\
E & Organ music            & O & Crow call \\
F & Classical guitar music & P & Glass shatter \\
G & Violin music           & Q & Car horn \\
H & Trumpet music          & R & Baby crying \\
I & Clarinet music         & S & Door slam \\
J & Alto saxophone music   & T & Alarm sound \\
\hline
\end{tabular}
\end{table}

The generated letter is mapped to its category using Table~\ref{tab:candidate_labels}. A prediction is correct when this category matches the specified listening target. Using a common set of answer codes keeps the scoring independent of how the model would otherwise name the sound.

\subsection{Free-Form Classification Prompt}

For free-form classification, we remove the candidate list and ask the model to describe the sound:

\begin{quote}
\small
\texttt{What is the main sound in this audio? Reply with a short sound description only.}
\end{quote}

We normalize the generated description and map it to one of the twenty categories to compute classification accuracy. The mapping uses fixed, case-insensitive keyword patterns that include category names and common alternatives, such as ``violin'' and ``fiddle.'' A response is assigned a category when exactly one category matches these patterns. Responses with no matching category or with multiple matching categories do not count as correct predictions. The patterns are applied after generation and are not included in the prompt.

This readout tests whether the enhanced audio can elicit a description of the specified sound without presenting the model with candidate names. It uses the same target categories as constrained classification, making the two language-output readouts directly comparable.

\Needspace{8\baselineskip}
\subsection{Prompt-Free Retrieval}

Retrieval diagnoses alignment of the audio representations with the specified target category, without a textual query or a generated answer. We average each test audio-token sequence over time, normalize the resulting vector, and compare it with a separately constructed clean reference gallery. Within each category, we average the five largest cosine similarities. The category with the highest average is the prediction.

More precisely, let $q$ denote the normalized mean of the query tokens and $r_{j,i}$ the normalized mean embedding of gallery recording $i$ from category $j$. If $\mathcal{I}_j(q)$ contains the indices of the five gallery recordings with the largest cosine similarities to $q$ within category $j$, the score and prediction are
\begin{equation}
s_j(q)=\frac{1}{5}\sum_{i\in\mathcal{I}_j(q)}q^{\top}r_{j,i},
\qquad
\widehat c=\underset{j}{\operatorname{argmax}}\;s_j(q).
\label{eq:retrieval_score}
\end{equation}
Each query therefore produces one category prediction from its similarity to clean audio examples. For LTM-AE, pooling is performed after enhancement, so this diagnostic measures the resulting representation before language generation.

The gallery is distinct from the long-term memory references and excludes overlapping audio windows. In the PCA conditions, the specified target category selects the long-term memory used to enhance the query tokens before retrieval. The retrieval stage receives these category-conditioned representations without a separate target cue.

\subsection{Evaluation Conditions}

Every model and evaluation protocol uses the same five conditions:

\begin{itemize}
    \item \textbf{Clean}: classification or retrieval using clean target audio.
    \item \textbf{Mixed}: evaluation on the unenhanced audio mixture, without the target category as input.
    \item \textbf{PCA ($N=20$)}: PCA enhancement using 20 clean memory examples per class.
    \item \textbf{PCA ($N=50$)}: PCA enhancement using 50 clean memory examples per class.
    \item \textbf{PCA ($N=100$)}: PCA enhancement using 100 clean memory examples per class.
\end{itemize}

All large-scale mixtures are evaluated under the same fixed $-10$ dB mixing condition. The clean condition uses the target recording corresponding to each mixture, and the enhanced conditions operate on that same mixture. The PCA conditions denote LTM-AE with long-term memory constructed from the indicated number of clean references.

For each readout, we first compute the fraction of correct predictions within each target category and then average across the twenty categories. Each category contributes 100 test recordings, so this macro-average also equals accuracy over all 2,000 test recordings. Gains are differences between enhanced and raw-mixture accuracies, reported in percentage points. The same scoring rule is retained across conditions within each readout.

\clearpage
\section{Additional Method Details}
\label{app:method_details}

We give the model interfaces and implementation details underlying LTM-AE, followed by the token decomposition that explains its action.
We then describe the learned gate used for speech transcription and compare the two solvers used to construct long-term memory.

\subsection{Audio Interfaces and Memory Construction}
\label{app:audio_interfaces}

The mapping $F_{\theta}$ includes the modules that convert audio into input embeddings for each ALLM's language model.
For \texttt{Qwen2-Audio}, we use the audio encoder outputs after the multimodal projector~\citep{chu2024qwen2}.
For \texttt{Kimi-Audio}, we use the fused embeddings formed from discrete semantic token embeddings and continuous Whisper features processed by the adaptor~\citep{ding2025kimi}.
For \texttt{Step-Audio-2-mini}, we use the audio adaptor outputs~\citep{wu2025step}.
Each model has a separate category long-term memory in its own input embedding space.
The clean reference recordings and incoming mixtures pass through the same mapping within each model, so the stored mean and directions can be applied directly to the mixture embeddings.
After enhancement, the refined embeddings occupy the original audio positions in the language model's input sequence.
The text token embeddings and sequence structure remain unchanged, and the language backbone receives no separate target cue.

Memory construction includes every valid audio token without temporal pooling or individual token normalization.
Each token contributes equally, so a reference clip contributes in proportion to its token count.
Keeping individual tokens allows the reference matrix to describe variation within a recording as well as differences across clean recordings.
The mean centers these observations, and the retained directions summarize their dominant variation in the same embedding coordinates.
Our default implementation uses \texttt{torch.pca\_lowrank} to obtain the retained directions of the centered reference matrix.
The alternative implementation computes them with \texttt{torch.linalg.svd}.
Both implementations construct the mean and directions used in Eq.~(\ref{eq:memory}).
The \texttt{Qwen2-Audio} implementation centers the token matrix explicitly and runs the randomized solver with \texttt{center=False}.
Its default settings compute 128 directions with five power iterations, after which validation selects how many directions to retain.
The full SVD implementation decomposes the same centered matrix and retains its leading right singular vectors.
The offline archive also stores singular values and construction metadata.
Inference uses the mean and selected basis columns.

Calibration reuses this stored mean and ordered basis while evaluating candidate ranks and interpolation weights on paired validation recordings.
For each candidate rank, the reconstruction uses the corresponding prefix of the basis matrix.
The enhanced mixture tokens are compared with their aligned clean target tokens using the mean squared error in Eq.~(\ref{eq:memory_selection}).
Thus, calibration determines how much of the stored variation to retain and how strongly to apply its correction to the current recording.
The selected settings are saved for each model and category together with the corresponding long-term memory.

The clean reference clips and paired validation recordings are no longer needed at inference.
For each token, enhancement consists of computing coefficients along the retained directions and mapping them back to the audio embedding space.
These two matrix products use $U_{c,k}$ directly, without forming the $d\times d$ matrix $P_c$.
For a sequence of $T$ tokens, the additional computation costs $O(Tdk_c)$ operations and the deployed memory requires $O(d(k_c+1))$ storage.
For fixed $d$ and $k_c$, the deployed memory size is independent of the number of reference clips $N$.
Increasing the reference count changes the observations used to estimate long-term memory, while inference continues to use its compact mean and retained directions.

\Needspace{12\baselineskip}
\subsection{Effect of Memory Guidance}
\label{app:memory_effect}

The interpolation rule preserves the component of a token described by the retained directions and scales the remaining component.
To make this operation explicit, substituting the memory reconstruction into the interpolation rule gives
\begin{equation}
\begin{aligned}
z'_t
&=(1-\alpha_c)z_t+\alpha_c\bigl[\mu_c+P_c(z_t-\mu_c)\bigr]\\
&=\mu_c+\bigl[(1-\alpha_c)I+\alpha_c P_c\bigr](z_t-\mu_c)\\
&=\mu_c+\bigl[P_c+(1-\alpha_c)(I-P_c)\bigr](z_t-\mu_c).
\end{aligned}
\label{eq:memory_decomposition}
\end{equation}
This yields Eq.~(\ref{eq:residual_shrinkage}) in the main text.
Since the columns of $U_{c,k}$ are orthonormal, $P_c$ and $I-P_c$ describe orthogonal components of the centered token.
The retained component has weight one throughout enhancement, whereas the orthogonal component has weight $1-\alpha_c$.
At $\alpha_c=0$, both components retain their original weights.
As $\alpha_c$ increases, the representation moves toward the reconstruction around the stored mean while preserving the coefficients obtained from the current token.

We can also express the same operation relative to the aligned clean target token used for calibration.
Let $e_t=z_t-z_t^{\mathrm{clean}}$ and $e'_t=z'_t-z_t^{\mathrm{clean}}$.
Using the same decomposition, we obtain
\begin{equation}
e'_t=P_c e_t+(1-\alpha_c)(I-P_c)e_t
-\alpha_c(I-P_c)(z_t^{\mathrm{clean}}-\mu_c).
\label{eq:memory_error}
\end{equation}
The final term vanishes when the memory reconstructs the clean target token exactly.
In that case, enhancement preserves the error along the retained directions and attenuates its orthogonal component.
For a general clean target, the final term records its component outside the retained directions.
The validation objective evaluates the complete error $e'_t$, including this term, when selecting the retained rank and interpolation weight.
This connects the decomposition to calibration on clean target embeddings and explains why the two settings are selected together.
This identity is expressed in the audio embedding space of the ALLM.

\subsection{Token-Level Gating for Speech Transcription}
\label{sec:memory_gate}

The speech transcription experiment in Section~\ref{sec:speech_transcription} extends LTM-AE with a learned scalar gate on the memory correction at each temporal position.
The speech category specifies the listening target, while the transcript depends on the content of the current recording.
The gate adapts the strength of long-term memory guidance at each position for this content recovery task.
This experiment trains a gate while keeping the category long-term memory and all ALLM parameters fixed.
Let $\delta_t=z'_t-z_t$ be the correction from the calibrated reconstruction.
The gate takes the concatenation of $z_t$ and $\delta_t$ as input.
It applies layer normalization followed by two linear layers with a hidden width of 256 and a GELU activation between them.
The final sigmoid produces a scalar weight for each token.
The same gate network is applied at every temporal position, with its output determined by that position's incoming token and memory correction.
With gate value $g_t\in[0,1]$, the resulting token is
\begin{equation}
z_t^{\mathrm{gate}}=z_t+g_t\delta_t
 =z_t+g_t\alpha_c(\widehat z_t-z_t).
\label{eq:memory_gate}
\end{equation}
The scalar applies to every feature coordinate at that position.
Because $\delta_t$ already includes $\alpha_c$, the effective reconstruction weight is $g_t\alpha_c$.
The gate can therefore reduce the calibrated correction for individual tokens.
At $g_t=0$, the original embedding is retained, while $g_t=1$ recovers the ungated LTM-AE embedding.
Intermediate values interpolate between these two representations without changing the number or order of tokens passed to the language backbone.

The gate is trained with a weighted combination of token reconstruction error and transcription loss,
\begin{equation}
\mathcal{L}(\phi)=
\lambda_{\mathrm{tok}}\operatorname{MSE}(Z^{\mathrm{gate}},Z^{\mathrm{clean}})
 +\lambda_{\mathrm{text}}\mathcal{L}_{\mathrm{CE}}
 (y\mid Z^{\mathrm{gate}}),
\label{eq:gate_objective}
\end{equation}
where $\phi$ denotes the gate parameters and $y$ is the reference transcript.
The token reconstruction term compares the gated embeddings with those of the aligned clean speech recording.
The transcription term connects the gate to the words decoded from these embeddings.
The transcription loss uses teacher forcing with prompt and padding positions excluded from supervision.
Gradients pass through the language backbone to the gate, while only $\phi$ is updated.
At inference, the gate computes its weights from the mixture embeddings and their memory corrections, with no clean target recording or reference transcript required.
Appendix~\ref{app:speech_results} gives the training and selection settings and the corresponding transcription results.

\subsection{PCA--SVD Solver Cross-Check}
\label{app:solver_check}

We validate the memory-construction solver by recomputing the \texttt{Qwen2-Audio} memories with a full singular value decomposition (SVD).
The SVD uses \texttt{torch.linalg.svd} with the \texttt{gesvd} driver on the same centered token matrices used by the randomized PCA implementation. For each of the twenty categories and each reference count
$N\in\{20,50,100\}$, the retained rank and interpolation weight are selected
on the validation partition only. The final evaluation contains 2,000
mixtures, and prompt-free retrieval averages the five largest cosine
similarities within each category.

\begin{table}[htbp]
\centering
\caption{Solver cross-check for Qwen2-Audio prompt-free retrieval.
Accuracies are macro averages over the twenty categories; the final column
reports the absolute difference in percentage points.}
\label{tab:solver_check}
\small
\begin{tabular}{lrrr}
\toprule
Memory size & Randomized PCA & Full SVD & Difference (pp) \\
\midrule
$N=20$  & 86.75 & 86.75 & 0.00 \\
$N=50$  & 86.75 & 86.75 & 0.00 \\
$N=100$ & 86.10 & 86.10 & 0.00 \\
\bottomrule
\end{tabular}
\end{table}

The two solvers select the same $(k,\alpha)$ configuration for all
sixty category--memory-size pairs, and their per-class retrieval accuracies
are identical.
This agreement supports using either solver to implement the same long-term memory construction objective.

\clearpage
\section{MultiPerception Experimental Results}
\label{app:additional_experiments}

We provide the construction details of MultiPerception and additional analyses of the diagnostic readouts.
The data combine multiple sound categories within each mixture, allowing us to examine enhancement toward a specified target in the presence of competing sources.
We then compare the evaluated reference counts and inspect representative language outputs to connect the aggregate gains with individual listening targets.

\subsection{Data Sources and Mixture Construction}
\label{app:data_sources}

MultiPerception combines the source collections in Table~\ref{tab:final20_sources} to cover musical instruments, human speech, animal vocalizations, and environmental events.
This range of categories supports evaluation of the same enhancement procedure on both speech and non-speech targets.
We standardize each clip to three seconds at 16\,kHz with one audio channel.
Each category has three source-group-disjoint partitions of 100 clips for memory construction, validation, and final evaluation.
Audio quality checks and hash-based duplicate removal are applied before evaluation.
The memory conditions use $N\in\{20,50,100\}$ clips from the memory partition.
These partitions provide 6,000 clean clips in total.
The validation and evaluation targets each retain their clean counterpart, so the same construction supports calibration against clean target tokens and comparison with clean-audio readouts.

\begin{table}[htbp]
\centering
\caption{Source collections used to construct the twenty target categories. Some categories combine several collections.}
\label{tab:final20_sources}
\footnotesize
\begin{tabular}{p{0.24\textwidth}p{0.68\textwidth}}
\toprule
Target category & Source collections \\
\midrule
Piano & MAESTRO~\citep{hawthorne2018enabling} \\
Dog vocalization & Dog play/pant~\citep{cuaya2026annotated}, FSD50K~\citep{fonseca2021fsd50k}, and UrbanSound8K~\citep{salamon2014dataset} \\
Human speech & LibriSpeech~\citep{panayotov2015librispeech} \\
Flute, organ & NSynth~\citep{engel2017neural} \\
Classical guitar & GAPS~\citep{riley2024gaps} \\
Violin, trumpet, clarinet, alto saxophone, drums & RWC Instruments Database~\citep{goto2003rwc} \\
Cat meow & CatMeows~\citep{ludovico2021catmeows} \\
Cow moo & CowRumIA~\citep{tenasalas2026cowrumia} \\
Frog croak & Glass Frog Calls~\citep{vargascastro2024glassfrog} \\
Crow call & CrowCall~\citep{thomas2025crowtools} and FSD50K~\citep{fonseca2021fsd50k} \\
Baby crying & Donate-a-Cry~\citep{veres2015donateacry} \\
Glass breaking, alarm & FSD50K~\citep{fonseca2021fsd50k} \\
Car horn & FSD50K~\citep{fonseca2021fsd50k} and UrbanSound8K~\citep{salamon2014dataset} \\
Door slam & DCASE 2013 isolated events~\citep{stowell2015detection} and FSD50K~\citep{fonseca2021fsd50k} \\
\bottomrule
\end{tabular}
\end{table}

For each validation or evaluation target, we choose three interferers from distinct non-target categories.
We first match each interferer's root mean square (RMS) amplitude to the target and then scale their sum to obtain a target-to-combined-interference ratio of $-10$\,dB.
The ratio is defined against the combined interference, so each mixture presents the target alongside three other sound sources under the same overall interference level.
Piano and dog targets use speech, flute, and organ as interferers.
Speech targets use dog, flute, and organ.
For each remaining target category, the interferer categories rotate through the other nineteen classes with a fixed seed.
This produces 2,000 validation mixtures and 2,000 evaluation mixtures, with the same evaluation audio used by all three ALLMs.
The mixture manifests record the target and interferer sources together with their gains and measured signal-to-noise ratio.
These records retain the correspondence between each mixture and the clean target used for calibration or evaluation.
The separate retrieval gallery contains 100 clean clips per category, with overlapping audio windows excluded as described in Appendix~\ref{app:prompts}.
It provides clean examples against which the readout measures alignment of the mixture representation with the specified target category.

\subsection{Results Across Memory Sizes and Categories}
\label{app:memory_size_figures}

\begin{figure}[htbp]
\centering
\includegraphics[width=\textwidth]{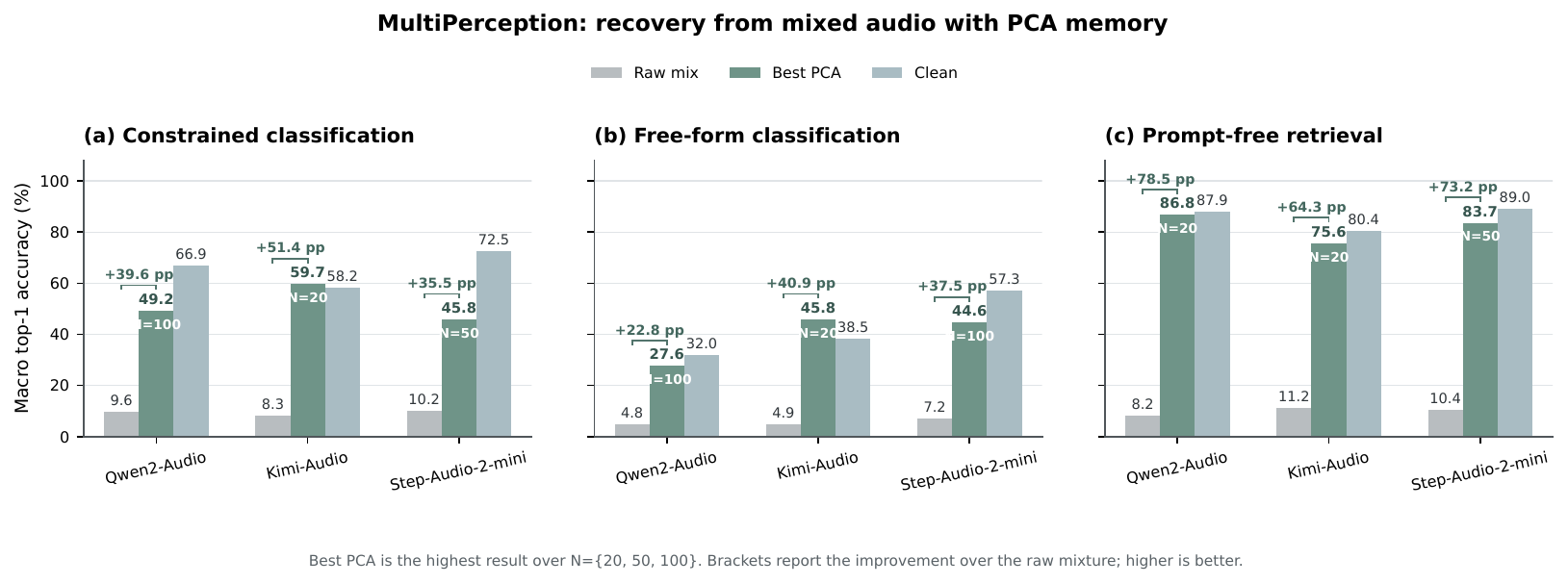}
\caption{MultiPerception results summarized by the highest macro accuracy over $N\in\{20,50,100\}$ for each model and protocol. ``Best PCA'' denotes this maximum among the evaluated LTM-AE memory sizes. The retained rank and interpolation weight are calibrated on validation data for each size. The numbers above brackets give gains over the raw mixture in percentage points.}
\label{fig:final20_best_memory}
\end{figure}

Figure~\ref{fig:final20_best_memory} summarizes the highest macro accuracy among the three evaluated reference counts for each model and protocol.
The displayed count is selected from the reported test results, so this summary differs from the fixed $N=20$ comparison in the main text.
Figure~\ref{fig:final20_classwise_gain_main} in the main text further shows
category-level variation.
Its maximum is taken separately within each category, rather than selecting one reference count for a whole model and protocol.
The complete values for each reference count appear in Appendix~\ref{app:detailed_results}.

The aggregate gains persist across all three reference counts for every model and readout.
For \texttt{Qwen2-Audio}, retrieval remains between 86.10\% and 86.75\%, compared with 8.25\% on raw mixtures, while constrained accuracy ranges from 47.20\% to 49.20\%.
For \texttt{Kimi-Audio}, free-form accuracy stays between 43.30\% and 45.80\%, compared with 4.90\% on raw mixtures.
For \texttt{Step-Audio-2-mini}, retrieval ranges from 78.50\% to 83.65\%, compared with 10.45\% on raw mixtures.
Thus, the improvement over raw mixtures is present throughout the evaluated range of reference counts.
The class-wise tables complement these aggregate comparisons by showing how individual sound categories respond to long-term memory guidance.

\Needspace{26\baselineskip}
\subsection{Representative Language-Output Cases}
\label{app:language_output_cases}

Table~\ref{tab:representative_language_cases} presents six representative
qualitative cases from the held-out MultiPerception evaluation set.
All enhanced outputs use the $N=20$ LTM-AE condition.
Free-form descriptions are mapped to the displayed categories using the
normalization rules in Appendix~\ref{app:prompts}.
The prompts remain unchanged between the raw and enhanced conditions and do not identify the selected target.

The cases illustrate how enhanced audio embeddings change which sound the language backbone reports.
In Case~2, \texttt{Qwen2-Audio} describes the competing dog sounds in the raw mixture but reports the specified alarm after enhancement.
Case~6 shows a corresponding change within a family of musical instruments, where \texttt{Step-Audio-2-mini} changes its description from a violin to the target flute.
The constrained outputs show the same pattern in a fixed answer space, including the frog croak in Case~5.
These examples make the diagnostic gains concrete by showing responses directed toward the specified sound under the same generic readout prompts.

\begin{table*}[htbp]
\centering
\caption{Representative language-output cases under acoustic interference.
The arrows for free-form classification indicate the normalized category
mapping.}
\label{tab:representative_language_cases}
\scriptsize
\setlength{\tabcolsep}{3pt}
\begin{tabular}{p{0.15\textwidth}p{0.38\textwidth}p{0.20\textwidth}p{0.20\textwidth}}
\toprule
Model and readout & Case description & Raw mixture output & LTM-AE output \\
\midrule

Qwen2-Audio, constrained
&
Case 1.
Target: flute music.
Interferers: clarinet music, alto saxophone music, concert drums.
&
\texttt{J} $\rightarrow$ alto saxophone music
&
\texttt{D} $\rightarrow$ flute music
\\[4pt]

Qwen2-Audio, free-form
&
Case 2.
Target: alarm sound.
Interferers: dog vocalization, human speech, flute.
&
``a dog barking and growling''
$\rightarrow$ dog vocalization
&
``alarm''
$\rightarrow$ alarm sound
\\[4pt]

Kimi-Audio, constrained
&
Case 3.
Target: classical guitar music.
Interferers: piano music, dog vocalization, human speech.
&
\texttt{D} $\rightarrow$ flute music
&
\texttt{F} $\rightarrow$ classical guitar music
\\[4pt]

Kimi-Audio, free-form
&
Case 4.
Target: human speech.
Interferers: dog vocalization, flute music, organ music.
&
``a man speaks and a dog barks''
$\rightarrow$ dog vocalization
&
``a man speaking with a beeping sound''
$\rightarrow$ human speech
\\[4pt]

Step-Audio-2-mini, constrained
&
Case 5.
Target: frog croak.
Interferers: alarm sound, piano music, dog vocalization.
&
\texttt{B} $\rightarrow$ dog vocalization
&
\texttt{N} $\rightarrow$ frog croak
\\[4pt]

Step-Audio-2-mini, free-form
&
Case 6.
Target: flute music.
Interferers: violin music, trumpet music, clarinet music.
&
``A violin is playing a single note.''
$\rightarrow$ violin music
&
``A flute is playing a long note.''
$\rightarrow$ flute music
\\

\bottomrule
\end{tabular}
\end{table*}

\subsection{Response to Long-Term Memory When the Target Is Absent}
\label{app:target_absence}

To examine whether selecting a long-term memory induces mentions of its category when the sound is absent, we apply piano long-term memory to 100 recordings without piano sounds.
We query \texttt{Qwen2-Audio} before and after enhancement with the same free-form prompt: ``What is the main sound in this audio? Reply with a short sound description only.''
We count how many responses mention piano.

\begin{table}[htbp]
\centering
\caption{Piano mentions in free-form responses to 100 recordings without piano sounds. Each entry gives the number of responses mentioning piano out of the 100 inputs.}
\label{tab:piano_absence}
\begin{tabular}{lcc}
\toprule
Model & Original audio & Piano LTM-AE \\
\midrule
Qwen2-Audio & 0/100 & 0/100 \\
\bottomrule
\end{tabular}
\end{table}

Table~\ref{tab:piano_absence} shows zero piano mentions under both conditions.
Selecting piano long-term memory did not induce the model to report piano in these recordings.

\clearpage
\section{Supplementary Speech Transcription Results}
\label{app:speech_results}

Speech transcription extends selective target perception from category responses to the content of the selected target. The listening cue identifies speech, while the transcript must be recovered from the current mixture. This experiment evaluates the gated extension described in Appendix~\ref{sec:memory_gate}, which adjusts the strength of long-term memory guidance at each audio token.

\subsection{Validation Selection and Final Evaluation}

The speech experiment uses \texttt{Qwen2-Audio} to transcribe English speech from mixtures.
The target recordings are three-second LibriSpeech segments sampled at 16~kHz~\citep{panayotov2015librispeech}.
The category long-term memory uses $N=500$ clean speech clips with $k=8$ and $\alpha=0.7$.
We train the gate in Appendix~\ref{sec:memory_gate} for four epochs with a learning rate of $10^{-3}$ and $\lambda_{\mathrm{text}}=1$.
The category long-term memory and all ALLM parameters remain unchanged.
The gate receives the original audio token together with its correction from LTM-AE and produces a scalar weight at each temporal position. It therefore adapts the contribution of the same speech long-term memory throughout the recording while preserving the sequence passed to the language backbone. The transcription loss supervises the words in the target recording, and the token reconstruction loss compares the gated embeddings with the corresponding clean target embeddings.

We divide 100 development mixtures into 80 training and 20 selection examples using seed 20260919.
For each $\lambda_{\mathrm{tok}}\in\{0.0,0.1,\ldots,1.0\}$, we train on the 80 examples and compute word error rate (WER) on the selection set.
The sweep varies the contribution of token reconstruction while keeping the transcription-loss weight and the speech long-term memory settings fixed.
Figure~\ref{fig:token_gate_selection} in the main text shows the resulting
loss-weight sweep.
Both $0.1$ and $0.8$ reach 20.57\% WER, and the predefined tie rule selects the smaller weight.
The second row of the figure subtracts this minimum from each selection WER and reports the difference in percentage points.
We then train the selected configuration on all 100 development examples and evaluate it on 100 held-out test mixtures.
The test set is not used to select the loss weight in this protocol.

WER is the total number of substitutions, deletions, and insertions divided by the number of reference words.
The selected configuration makes 36 errors on 175 selection words and 130 errors on 880 test words.
The final test result is 14.77\% WER, comprising 63 substitutions, 48 deletions, and 19 insertions.
These counts are summed across the test recordings before division by the total number of reference words, following Eq.~(\ref{eq:wer}). Substitutions count incorrectly transcribed words, deletions count omitted reference words, and insertions count additional words in the prediction.

Table~\ref{tab:speech_wer_formal} compares this result with the raw-mixture WER of 23.07\% and the ungated LTM-AE WER of 33.75\% on the matched test set. The selected gate reduces WER by 8.30 percentage points relative to raw mixtures. With the speech category and long-term memory fixed, this improvement concerns the words spoken in each recording. The result supports using token-level control to extend long-term memory guidance to target content recovery.

\FloatBarrier

\clearpage

\end{document}